# Simple Formulas for Output Interception Power Estimation of Uni-Traveling Carrier Photodiodes

Keye Sun, Junyi Gao, Yuan Yuan, Joe. C. Campbell, and Andreas. Beling

*Abstract*—Simple analytical expressions for estimation of second order output intercept point (OIP2) and third order output intercept point (OIP3) of surface normal uni-traveling carrier (UTC) and modified uni-traveling carrier (MUTC) photodiode (PD) are derived. These equations are valuable for estimation of OIP for high power (M)UTC-PDs during the design phase.

## I. INTRODUCTION

MICROWAVE photonics, bridging the photonics and microwave engineering, attracts lots of interest over the past decades both in research community and commercial companies. Microwave photonics utilizes the advantages of photonics to achieve functionalities at microwave frequency range, which is difficult using electronics [1, 2].

PD is one of the most critical components of a microwave photonic systems. Usually in these systems, the operation frequencies are high aiming for fast speed. Moreover, strong optical power is desired for better system performance such as signal to noise ratio and spurious-free dynamic range, etc [3-5]. These two facts put stringent requirements on PDs used in microwave photonic systems. PDs need to maintain high bandwidth under high photocurrent operation. MUTC PD is one of the designs for high-power microwave PDs. In such structures, fast electrons are drifted across the depletion region while slow holes relax within the short dielectric relaxation time. [6]. To date, surface normal MUTC PDs with various designs have demonstrated high output RF power from tens of GHz to over 100 GHz [7-10].

Under high-power operation conditions, PDs exhibit nonlinear behavior and generate high-order harmonic outputs. The nonlinearity of the PDs affects the system performance in many applications such as communications [11], analog links [3], microwave signal generation [12], etc. As a result, the linearity of MUTC PDs needs to be carefully studied. The nonlinearity of the MUTC PDs can be characterized by OIP2 an OIP3 parameters. Unlike the bandwidth and saturation, it is difficult to estimate the OIP during the epi-layer structure design phase. Previously, the OIP of the MUTC-PD is carefully investigated [13-16]. However, measurement results of a fabricated PD such as the responsivity, voltage- and current-dependent junction capacitance are needed for the analysis of OIP of the PDs. It is desirable to have some rough estimation on OIP in early design phase just based on epi-layer structure without resorting to all the measured parameters beforehand. As far as we know, such analytical formulas are not available in literature. In this work, simple analytical formulas as well as numerical methods for OIP estimation for (M)UTC PDs are provided.

## II. THEORY

### A. Equivalent Circuit Model and Harmonics Calculation

As shown in [14] that at low frequency, the nonlinearity of the MUTC-PD is dominated by voltage-dependent responsivity caused by Franz-Keldysh effect. However, the nonlinearity of the junction capacitance starts to dominate the nonlinear behavior at relatively high frequency since more photocurrent is shunt through it. In this model, the nonlinearity resulted from junction capacitance is calculated while Franz-Keldysh effect is neglected aiming at high frequency operation.

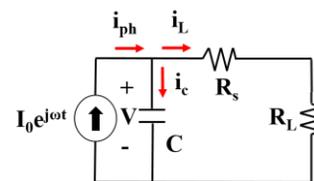

Fig. 1. Circuit model used to calculate nonlinearity of a MUTC-PD.

A simple circuit model of a PD [17] was used as shown in Fig. 1 where $C$, $R_s$ and $R_L$ denote the junction capacitance, series resistance and load resistance of the PD. Assuming a single tone optical signal at $\omega$ is incident on the PD, the photocurrent $i_{ph}$ of $I_0 e^{j\omega t}$ is generated. The photocurrent splits into two portions of which $i_L$ flows to the load and $i_c$ flows through the junction capacitance. At high photocurrent, the junction capacitance $C$ is no longer a constant and becomes modulated due to the following two facts. Firstly, the voltage drop on the load modulates the depletion width and thus the junction capacitance $C$. Secondly, the photocurrent $i_{ph}$ changes the electric-field profile inside the PD and the depletion width varies under the condition that the voltage across the PD is fixed. Combining the

The authors are with the Department of Electrical and Computer Engineering, University of Virginia, Charlottesville, VA-22903 USA (e-mail: ks2kz@virginia.edu, jg4fv@virginia.edu, yy6mf@virginia.edu, jcc7s@virginia.edu, ab3pj@virginia.edu).

two facts, the junction capacitance of the PD is a function of both voltage across the junction $V$ and photocurrent $i_{ph}$. For the first order approximation, C can be expressed as

$$C = C_0 + \frac{\partial C}{\partial V}\Delta V + \frac{\partial C}{\partial i}\Delta i_{ph} = C_0 + \frac{\partial C}{\partial V}i_L R + \frac{\partial C}{\partial i}\Delta i_{ph} \quad (1)$$

where $R=R_s+R_L$. Thus the current nodal equation can be written as

$$i_{ph} = i_c + i_L = \frac{d}{dt}[C(V, i_{ph}) \cdot V] + i_L \quad (2)$$

Plug in $i_{ph}$, C into the nodal equation,

$$i_L + R\frac{d}{dt}\left[\left(C_0 + I_0 e^{j\omega t}\frac{\partial C}{\partial i}\right)i_L + i_L^2 R\frac{\partial C}{\partial V}\right] = I_0 e^{j\omega t} \quad (3)$$

The output current can have high order harmonics due to the modulation on the junction capacitance. Neglecting harmonics higher than the 3rd order, the load current can be written as

$$i_L = i_1 e^{j\omega t} + i_2 e^{j2\omega t} + i_3 e^{j3\omega t} + \cdots \quad (4)$$

where $i_1$, $i_2$ and $i_3$ are the amplitudes of the fundamental, 2nd and 3rd order harmonics. Plug (4) into (3) and match each frequency component,

$$i_1 = \frac{I_0}{1+j\omega RC_0} \quad (5)$$

$$i_2 = -2j\omega \frac{i_1^2 R^2 \frac{\partial C}{\partial V} + i_1 I_0 R \frac{\partial C}{\partial i}}{1+2j\omega RC_0} \quad (6)$$

$$i_3 = -3j\omega \frac{2i_1 i_2 R^2 \frac{\partial C}{\partial V} + i_2 I_0 R \frac{\partial C}{\partial i}}{1+3j\omega RC_0} \quad (7)$$

Then the OIP2 and OIP3 can be calculated based on these harmonics. The amplitude of the three harmonics are expressed in functions of $\frac{\partial C}{\partial V}$ and $\frac{\partial C}{\partial i}$. Next, the junction capacitance are calculated as a function of junction voltage $V$ and photocurrent $i_{ph}$.

B. *Calculation of $\frac{\partial C}{\partial V}$*

The band diagram of a UTC is shown in Fig. 2 (a) and corresponding electric field profile is shown in Fig. 2 (b). The layers are assumed to be fully depleted under bias voltage $V$. The n-contact layer is heavily-doped and the depletion width inside it is negligible. The p-type absorber has a doping level of $\rho$ and hence the depletion width $W_p$ can vary as a function of voltage across the junction. The drift layer is assumed to be intrinsic. The dielectric constant of absorber and drift layer are $\epsilon_1$ and $\epsilon_2$ respectively.

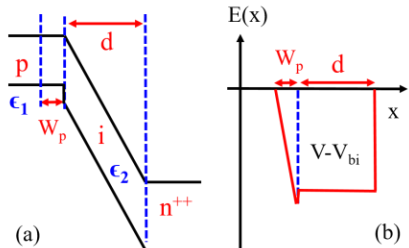

Fig. 2. (a) Band-diagram of a UTC-PD. (b) electric field profile in the PD.

Since the applied voltage across the PD is $V$, the total area under the electric field curve is equal to $V$-$V_{bi}$ where $V_{bi}$ is the built-in voltage.

$$\frac{W_p^2}{2} \cdot \frac{\rho}{\epsilon_1} + W_p d \frac{\rho}{\epsilon_2} = V - V_{bi} \quad (8)$$

Thus

$$\frac{\partial W_p}{\partial V} = \frac{1}{\frac{\rho W_p}{\epsilon_1} + \frac{\rho d}{\epsilon_2}} \quad (9)$$

The junction capacitance is a series connection of the capacitor due to the depleted portion in the absorber and the capacitor due to the drift layer. As a result, the derivative of junction capacitance relative to the voltage is

$$\frac{\partial C}{\partial V} = \frac{\partial}{\partial V}\left(\frac{1}{\frac{W_p}{\epsilon_1 A} + \frac{d}{\epsilon_2 A}}\right) = \frac{-C^3}{\rho \epsilon_1 A^2} \quad (10)$$

where A is the PD area.

C. *Calculation of $\frac{\partial C}{\partial i}$*

*C.1 UTC-PD*

The band diagram of a UTC is shown in Fig. 2 (a). The electric field under dark and light illumination condition are shown in Fig. 3 (b). Under light illumination, electrons are populated in the drift layer and result in a negative slope in the electric field profile. In order to keep a constant voltage drop across the device, the depletion width in the absorber should reduce accordingly. As a result, the junction capacitance varies as a function of photocurrent.

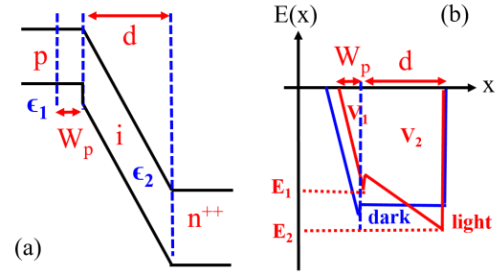

Fig. 3. (a) Band-diagram of a UTC-PD. (b) electric field profile in the PD.

Using Poisson equation, the electric field is calculated to be

$$E_1(W_p^-) = -\frac{\rho W_p}{\epsilon_1} \quad (11)$$
$$E_2 = E_1(W_p^+) - \frac{J}{v_{e2}\epsilon_2} \quad (12)$$

where $J$ is the photocurrent density and $v_{e2}$ is the carrier velocity. Saturated electron velocity is assumed inside the drift layer. The displacement vector is continuous at the interface between the absorber and drift layer

$$\epsilon_1 E_1(W_p^-) = \epsilon_2 E_1(W_p^+) \quad (13)$$

The area under the curve in Fig. 3 (b) should be equal to $V - V_{bi}$.

$$V_1 + V_2 = V - V_{bi} \quad (14)$$

where $V_1 = \frac{W_p^2}{2}\frac{\rho}{\epsilon_1}$ and $V_2 = \left(E_2(W_p^+) + E_2\right)\frac{d}{2}$. Using (12) ~ (14),

$$\frac{\partial W_p}{\partial J} = \frac{-A\gamma_{UTC}}{W_p/\epsilon_1 + d/\epsilon_2} \quad (15)$$

where $\gamma_{UTC} = \frac{d^2}{2v_2\epsilon_2\rho}$.

Thus, the derivative of junction capacitance relative to photocurrent is calculated to be

$$\frac{\partial C}{\partial i} = C^3 \cdot \frac{\gamma_{UTC}}{\epsilon_1 A^3} \quad (16)$$

*C.2 MUTC-PD*

For MUTC-PD, the derivation is the same as UTC-PD. However, the photo-generated current distribution inside depleted absorber should be considered since the electric field profile is affected. Assuming negligible recombination and exponential photo-generation rate under steady state, the photocurrent density is calculated as,

$$J_e = \frac{-J}{e^{\alpha W_{ad}} - e^{-\alpha W_{an}}}(e^{\alpha x} - e^{-\alpha W_{an}}) \quad (17)$$
$$J_h = \frac{-J}{e^{\alpha W_{ad}} - e^{-\alpha W_{an}}}(e^{\alpha W_{ad}} - e^{\alpha x}) \quad (18)$$

where $J_e$ and $J_h$ are electron, hole and total current density, respectively. $J_e$ and $J_h$ are negative since they flow in the negative direction. $J$ is the total photocurrent density and is set to be a positive number. $W$ is the total thickness of the absorber. $\alpha$ is the optical absorption coefficient of the absorber layer. The band-diagram, current density and electric field profile are shown in Fig. 4 (a), (b) and (c). The curvature of electric field profile under light illumination inside the depleted absorber is resulted from the current density distribution where the holes are accumulated on the left and electrons are accumulated on the right. The profile on Fig. 4 (c) is just an illustration and the actual curvature is determined by the electron and hole density calculated using Poisson equation.

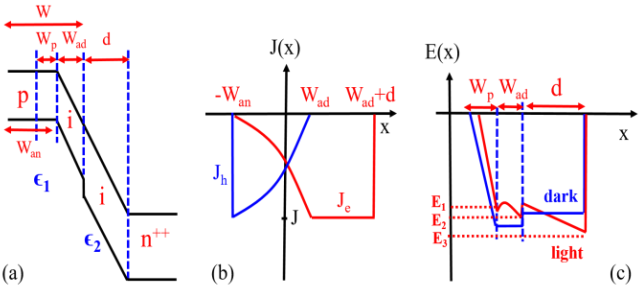

Fig. 4. (a) Band-diagram of a UTC-PD. (b) current density distribution (c) electric field profile in the PD.

Using Poisson equation and continuity of displacement vector at the interface between absorber and depletion layer, the electric field profile can be derived

$$E(x) = \begin{cases} -\dfrac{\rho x}{\epsilon_1} & \text{for } 0 < x < W_p \\ -\dfrac{W_p\rho}{\epsilon_1} - \dfrac{J}{(e^{\alpha W_{ad}} - e^{-\alpha W_{an}})\epsilon_1} \cdot \left[\left(\dfrac{1}{V_{e1}} + \dfrac{1}{V_{h1}}\right)\dfrac{1-e^{\alpha x}}{\alpha} + \left(\dfrac{e^{\alpha W_{ad}}}{V_{h1}} + \dfrac{e^{-\alpha W_{an}}}{V_{e1}}\right)x\right] & \text{for } W_p < x < W_{ad} \\ -\dfrac{W_p\rho}{\epsilon_2} - \dfrac{J}{(e^{\alpha W_{ad}} - e^{-\alpha W_{an}})\epsilon_2} \cdot \left[\left(\dfrac{1}{V_{e1}} + \dfrac{1}{V_{h1}}\right)\cdot\dfrac{1-e^{\alpha W_{ad}}}{\alpha} + \left(\dfrac{e^{\alpha W_{ad}}}{V_{h1}} + \dfrac{e^{-\alpha W_{an}}}{V_{e1}}\right)W_{ad}\right] - \dfrac{Jx}{v_{e2}\epsilon_2} & \text{for } W_{ad} < x < W_{ad}+d \end{cases}$$

(19)

where $v_{e1}$, $v_{h1}$, $v_{e2}$ and $v_{h2}$ are electron and hole velocities in the absorber and drift layer, respectively. The area under the electric field curve is equal to the voltage applied across the device, $V - V_{bi}$. Pluging in (19) and the derivative of $W_p$ with respective to $J$ can be derived as

$$\frac{\partial W_p}{\partial J} = \frac{-A\gamma_{MUTC}}{(W_p + W_{ad})/\epsilon_1 + d/\epsilon_2} \quad (20)$$

where

$$\gamma_{MUTC} = \frac{1}{\rho(e^{-\alpha W_{an}} - e^{\alpha W_{ad}})} \cdot \left[-\frac{\left(1/\epsilon_1\alpha + W_d/\epsilon_2\right)(e^{\alpha W_{ad}}-1) - \frac{W_{ad}}{\epsilon_1}}{\alpha}\cdot\left(\frac{1}{v_{e1}}+\frac{1}{v_{h1}}\right) + \left(\frac{e^{\alpha W_{ad}}}{V_{h1}}+\frac{e^{-\alpha W_{an}}}{V_{e1}}\right)\cdot\left(\frac{W_{ad}^2}{2\epsilon_1}+\frac{W_{ad}W_d}{2\epsilon_2}\right)\right] + \frac{W_d^2}{\rho\epsilon_2 v_{e2}}$$

(21)

Finally,

$$\frac{\partial C}{\partial i} = C^3 \cdot \frac{\gamma_{MUTC}}{\epsilon_1 A^3} \quad (22)$$

*D. Summary*

The summary of the equations for fundamental, 2nd and 3rd harmonic tones calculation are listed below:

$$\text{Fundamental tone: } i_1 = \frac{I_0}{1+j\omega RC_0} \quad (23)$$

$$\text{2nd harmonic tone: } i_2 = -2j\omega\frac{i_1^2 R^2\frac{\partial C}{\partial V} + i_1 I_0 R\frac{\partial C}{\partial i}}{1+2j\omega RC_0} \quad (24)$$

$$\text{3rd harmonic tone: } i_3 = -3j\omega\frac{2i_1 i_2 R^2\frac{\partial C}{\partial V} + i_2 I_0 R\frac{\partial C}{\partial i}}{1+3j\omega RC_0} \quad (25)$$

where

$$\frac{\partial C}{\partial V} = \frac{-C^3}{\rho\epsilon_1 A^2} \quad (26)$$

$$\frac{\partial C}{\partial i} = \begin{cases} C^3 \cdot \dfrac{\gamma_{UTC}}{\epsilon_1 A^3} & \text{for UTC} - \text{PD} \\ C^3 \cdot \dfrac{\gamma_{MUTC}}{\epsilon_1 A^3} & \text{for MUTC} - \text{PD} \end{cases} \quad (27)$$

$$\gamma_{UTC} = \frac{d^2}{2v_2\epsilon_2\rho} \quad (28)$$

$$\gamma_{MUTC} = \frac{1}{\rho(e^{-\alpha W_{an}} - e^{\alpha W_{ad}})}$$
$$\cdot \left[ \frac{\frac{W_{ad}}{\epsilon_1} - \left(\frac{1}{\epsilon_1\alpha} + \frac{W_d}{\epsilon_2}\right)(e^{\alpha W_{ad}} - 1) - \frac{W_{ad}}{\epsilon_1} \cdot \left(\frac{1}{v_{e1}} + \frac{1}{v_{h1}}\right)}{\alpha} + \left(\frac{e^{\alpha W_{ad}}}{V_{h1}} + \frac{e^{-\alpha W_{an}}}{V_{e1}}\right) \cdot \left(\frac{W_{ad}^2}{2\epsilon_1} + \frac{W_{ad}W_d}{2\epsilon_2}\right) \right] + \frac{W_d^2}{2\rho\epsilon_2v_{e2}}$$
(29)

### III. NUMERICAL MODEL

In real situation, the carrier velocities are functions of electric field. Analytical formulas for OIP cannot be given in this case. A simple numerical model was resorted to for more accurate calculation of the OIP. The electric field-dependent carrier velocities are given below [18]:

$$v_e = \frac{\mu_e E + v_e^{sat}\beta E^\gamma}{1 + \beta E^\gamma}$$

$$v_h = v_h^{sat}\tanh(\frac{\mu_h E}{v_h^{sat}})$$

where $\mu_e$, $v_e^{sat}$, $\mu_h$, $v_h^{sat}$, $\beta$, and $\gamma$ are empirical fitting parameters. $E$ is the electric field.

The critical step is to find the depletion width $W_p$ inside the undepleted absorber. The numerical algorithm adjusts $W_p$ iteratively using scant method [19] until bias voltage condition is satisfied. The algorithm is shown in Fig. 5.

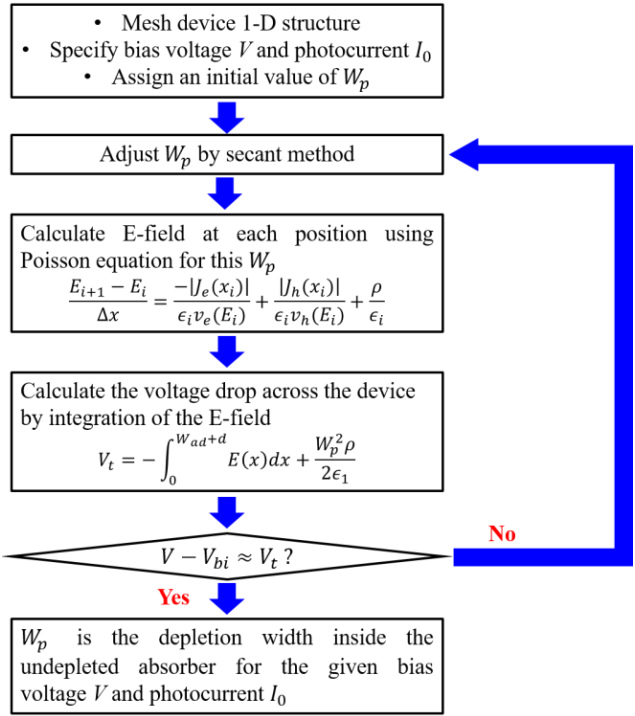

Fig. 5. Numerical algorithm for calculating $W_p$.

$\frac{\partial c}{\partial V}$ can be calculated by two $W_p$ values at two close bias voltage around the operation voltage $V$. $\frac{\partial c}{\partial i}$ can also be calculated in the same way. By plugging $\frac{\partial c}{\partial V}$ and $\frac{\partial c}{\partial i}$ into equation (23) ~ (25), OIP values can be calculated.

### IV. EXPERIMENTAL

The theory is based on 1-tone OIP measurement. In reality, 2- or 3-tone OIP measurement were used in experiments to avoid harmonic signals generated by the system in lieu of PD. In order to compare the theory and experimental results from the literature, conversion between 1-tone and 2- or 3-tone results should be included. The conversion is calculated as below.

Assuming the system has a nonlinear response of
$$i_{ph} = \alpha_1 P_{opt} + \alpha_2 P_{opt}^2 + \alpha_3 P_{opt}^3 + \cdots$$
where $\alpha_i$ is the $i$th nonlinear coefficient of the output signal and $P_{opt}$ is the optical input signal.

In 1-tone measurement, assuming input fundamental tone signal is $\cos\omega t$, the signal at the 3rd order harmonic frequency $3\omega$ is $\frac{\alpha_3}{4}\cos(3\omega t)$.

In 2-tone measurement, assuming input fundamental tone signal is $\cos\omega_1 t + \cos\omega_2 t$, the total signals at the 3rd order harmonic frequency of $2\omega_1 - \omega_2$ is $\frac{3\alpha_3}{4}\cos((2\omega_1 - \omega_2)t)$.

In 3-tone measurement, assuming input fundamental tone signal is $\cos\omega_1 t + \cos\omega_2 t + \cos\omega_3 t$, the total signals at the 3rd order harmonic frequency of $\omega_1 + \omega_2 - \omega_3$ is $\frac{3\alpha_3}{2}\cos((\omega_1 + \omega_2 - \omega_3)t)$.

From the above arguments, the OIP3 form the 1-, 2- and 3-tone measurement should result in the conversion relationship below

$$OIP3_{2-tone} = OIP3_{1-tone} - 4.8\ dB$$
$$OIP3_{3-tone} = OIP3_{1-tone} - 7.8\ dB$$
(30)

Experimental data from literature were fitted using both analytical formulas and numerical models. For numerical calculation, the parameters used are listed in table I.

TABLE I
MATERIAL PROPERTIES USED IN CALCULATION [18]

| Parameter | InP | InGaAs |
|---|---|---|
| $\epsilon_r$ | 12.5 | 13.8 |
| $v_e^{sat}$ | 0.85×10⁷ cm/s | 0.65×10⁷ cm/s |
| $v_h^{sat}$ | - | 0.48×10⁷ cm/s |
| $\mu_e$ | 3500 cm²/V·s | 8000 cm²/V·s |
| $\mu_h$ | 150 cm²/V·s | 300 cm²/V·s |
| $\beta$ | 7.4×10⁻¹³ | 7.4×10⁻¹⁰ |
| $\gamma$ | 3 | 2.5 |

The fitting results are listed in the table II. The conversion between 1-, 2- and 3-tone has been included using equation (30). As can be seen that the analytical results are higher than the literature ones. Numerical results agree with the literature ones reasonably well. The numerical and analytical methods are based on the same underlying physics except that the electric-field dependent carrier velocities are included in the numerical algorithm. It is clear that the variation of the carrier velocity due to the electric-field inside the PD is a main resource for intermodulation (IMD) nonlinearity.

TABLE II
COMPARISON RESULTS FROM LITERATURE

| Reference | IMD | Reported data (dBm) | Analytical results (dBm) | Numerical results (dBm) |
|---|---|---|---|---|
| [20] | OIP3 | 42 | 53 | 27 |
| [21] | OIP3 | 35 | 56 | 26 |
| [22] | OIP3 | 30 | 51 | 24 |
|  | OIP3 | 25 | 56 | 28 |
|  | OIP3 | 20 | 65 | 35 |
| [23] | OIP3 | 40 | 61 | 28 |
| [24] | OIP3 | 33 | 61 | 36 |

## V. Discussion

Both $\frac{\partial C}{\partial V}$ and $\frac{\partial C}{\partial i}$ are inversely proportional to the doping level $\rho$ in undepleted absorber. This is intuitive since the higher the doping level, the more difficult it is to vary the depletion width $W_p$. As a result, the junction capacitance is less sensitive to the variation of voltage and current, which results in higher OIP. Thus, increasing the doping level in undepleted absorber is one method to increase the OIP [16]. However, high doping level reduces the carrier life time and thus the quantum efficiency. There is a trade-off between the two.

The junction capacitance changes by varying the depletion width. Due to the inverse proportion relationship between junction capacitance and depletion width, C is more sensitive to depletion width variation when the depletion width is small and vice versa. As a result, both $\frac{\partial C}{\partial V}$ and $\frac{\partial C}{\partial i}$ increase with increasing capacitance.

The voltage dependence of junction capacitance is shown in equation (26). $\frac{\partial C}{\partial V}$ is negative since higher voltage leads to wider depletion region and thus less capacitance, which agrees with the intuition.

The current dependence of junction capacitance for UTC-PD is shown in equation (27) and (28). Note that $\frac{\partial C}{\partial i}$ in this case is positive. Increased photocurrent shrinks the depletion width in the undepleted absorber due to the more tilted electric field profile as shown in Fig. 3 (b) and thus increases the junction capacitance. From equation (14), the depletion width in the undepleted absorber can be derived

$$W_p = -d + d\sqrt{1 + (\frac{2\epsilon_2 v_{e2}}{\rho d^2} - \frac{J}{v_{e2}\rho})} \quad (31)$$

Since $W_p$ is a positive number, a maximum photocurrent can be calculated

$$J_{max} = \frac{2v_{e2}\epsilon_2(V - V_{bi})}{d^2} \quad (32)$$

which corresponds to the situation where the electric field completely collapses at the entrance of the drift layer.

$\frac{\partial C}{\partial i}$ for MUTC-PD is shown in equation (29) and it is more complicated than UTC-PD. It can be both positive and negative depending on the specific design of the epi-layers and carrier accumulation situation inside each layer.

The depleted absorber and drift layer are assumed to be intrinsic for the analytical formulas in order to simplify the derivation. In real PDs, these two layers are usually charge-compensation doped to increase the power handling capability. However, these results are still valid if doping is included since the doping only adds terms which are independent of current density J. Thus $\frac{\partial W_p}{\partial J}$ is not affected and neither is $\frac{\partial C}{\partial i}$.

Finally, the model reported here only takes into account the effect of space-charge effect on OIP, which has the major contribution at high frequency. It only provides a rough estimation on OIP during the early epi-layer design phase. For an accurate estimation, other facts should be taken into account such as bias-dependent responsivity [14], carrier diffusion, illumination condition, etc [25]. More sophisticated numerical simulation should be resorted to.

## VI. Conclusion

Analytic formulas and numerical model are given for rough estimation of OIP based on epi-layer structures. The limitation of this model are discussed. This model provides a valuable guidance for OIP estimation during the early epi-layer design of (M)UTC-PDs.